\documentclass{emulateapj}
\usepackage{latexsym}
\usepackage{dcolumn}
\usepackage{bm}
\input{epsf}
\newcommand{\be}{\begin{equation}}
\newcommand{\ee}{\end{equation}}
\newcommand{\ba}{\begin{eqnarray}}
\newcommand{\ea}{\end{eqnarray}}
\newcommand{\bfi}{\begin{figure}[t]
\epsfxsize=9cm
\epsffile}
\newcommand{\efi}{\end{figure}}
\newcommand{\apjl}{ApJ}
\newcommand{\apjs}{ApJS}
\newcommand{\aj}{AJ}

\begin{document}
\title{Dimming of supernovae and gamma ray busts by Compton Scattering
  and its cosmological implications }   
\author{Pengjie Zhang}
\affil{Shanghai Astronomical Observatory, Chinese Academy of
  Science, 80 Nandan Road, Shanghai, China, 200030}
\affil{Joint Institute for Galaxy and Cosmology (JOINGC) of
SHAO and USTC}
\begin{abstract}
Free electrons deplete photons from type
Ia supernovae through the (inverse) Compton scattering.  This Compton
dimming increases with redshift and reaches $0.004$ mag at $z=1$ and
$0.01$ mag at $z=2$. Although far from 
sufficient to invalidate the existence of dark energy, it can bias
constraint on dark energy at a level non-negligible for future
supernova surveys.  This effect is correctable and should be
incorporated in supernova analysis.  The Compton dimming has similar
impact on cosmology based on  gamma ray bursts as standard candles. 
\end{abstract}
\maketitle
\section{Compton dimming of Type Ia supernovae}
Type Ia supernovae (SNe Ia) are standardizable as  {\it
cosmological standard candles} to measure cosmological distance and
thus infer the expansion history of the universe. Current observations
on SNe Ia  
have enabled the discovery of the late time acceleration of the
universe \citep{Riess98,Perlmutter99}.  This discovery has
profound impact on fundamental physics, leading to either a
dominant dark  energy component with equation of state 
$w\equiv P/\rho<-1/3$ or significant deviations from general
relativity at around Hubble scale. Ongoing and planned supernova surveys
have the power to significantly improve these cosmological constraints
and hopefully clarify the role of the 
cosmological constant in our universe \citep{DETF}.

Various astrophysical processes, besides the possible intrinsic 
evolution in SN luminosity, can  degrade 
the standard candle merit of SNe Ia by altering the supernova flux.
An incomplete list include  
gravitational lensing magnification \citep{Holz98}, peculiar velocity
\citep{Hui06}, dust extinction 
and  incomplete K-correction. If not handled correctly, they can not only
increase statistical errors, but also systematically bias the
cosmological constraints. In this paper, we point out a new
source of systematical errors, relevant for precision cosmology. 

The universe is (almost) completely ionized after $z=6$. Free
electrons in the universe treat all low energy photons ($h\nu\ll
m_ec^2$) equally and Compton scatter\footnote{Free electrons have
  thermal and kinetic motions. So what happened actually is the
  inverse Compton 
  scattering. Inverse Compton scattering results in photon energy
  change at a level of $10^{-3}$ ($k_BT_e/m_ec^2$, $v/c\sim
  10^{-3}$). For CMB photons, this process results in the well known
  Sunyaev-Zel'dovich effect. However, this photon energy change is
  irrelevant for this paper since 
  none of scattered supernova photons reach us in time. } off them with equal
probability $\exp(-\tau)$, where $\tau$ is the 
Thomson optical depth
\be
\label{eqn:tau}
\tau(z)=\int_0^z \sigma_T
n^{free}_e(z)\frac{acdz}{H(z)}=\int_0^z\sigma_Tn_e(0)X_e(z)\frac{(1+z)^2}{H(z)}dz
\ .
\ee
Here, $\sigma_T$ is the Thomson cross section, $n^{free}(z)$ is the
number density of free electrons and 
$X_e(z)$ is the ionization fraction. $X_e(z)\simeq 1$ at $z<6$.
$H(z)$ is the Hubble constant and 
$a=1/(1+z)$ is the scale factor.  There are two competing effects on
the flux of 
a given celestial object. (A) For photons originally emitted towards
us, on the average $1-\exp(-\tau)\simeq \tau$ of 
them are Compton scattered away and escape of observation. (B) Photons
otherwise can not reach us may be scattered toward  us. If the 
celestial object is  non-evolving and the universe is static, these
two effects cancel each other exactly and the flux is
unchanged.\footnote{Tiny energy change in scattered photons is
 neglected in this statement.} This 
can be proved straightforwardly by the aid of photon number
conservation. However, none of the   
conditions is realistic. First of all, our universe is
expanding. Scattered photons take longer time to reach us and thus
suffer more energy loss. More importantly, SNe
Ia only last for months, much shorter than the time it
takes for photons to reach us. Scattered photons travel extra distance
and take extra time to reach us.  As a  consequence, only those photons 
originally emitted within a solid angle $\Omega_{scatter}\sim ct/D$
toward us can be scattered while reach us during the event
period. Here, $t$ and $D$ 
are the life time and the distance of SNe Ia,
respectively. For SNe Ia, 
$\Omega_{scatter}/4\pi\sim 10^{-11}\ll 1$. Thus it is virtually exact
that no scattered photons can reach us. Since effect B vanishes, for
SNe Ia,  Compton scattering alters the flux $F$ to $F\exp(-\tau)$. We call it
the Compton dimming. Its amplitude is 
\be
 \frac{\delta F}{F}=-2\frac{\delta D_L}{D_L}=\exp(-\tau)-1\simeq
 -\tau\ .
\ee
Here $D_L$ is the luminosity distance. This results in a systematic
shift in  the distance moduli $\mu$,  
\be
\Delta \mu=5\log_{10}(1+\delta D_L/D_L)\simeq 1.086\tau\ .
\ee
\bfi{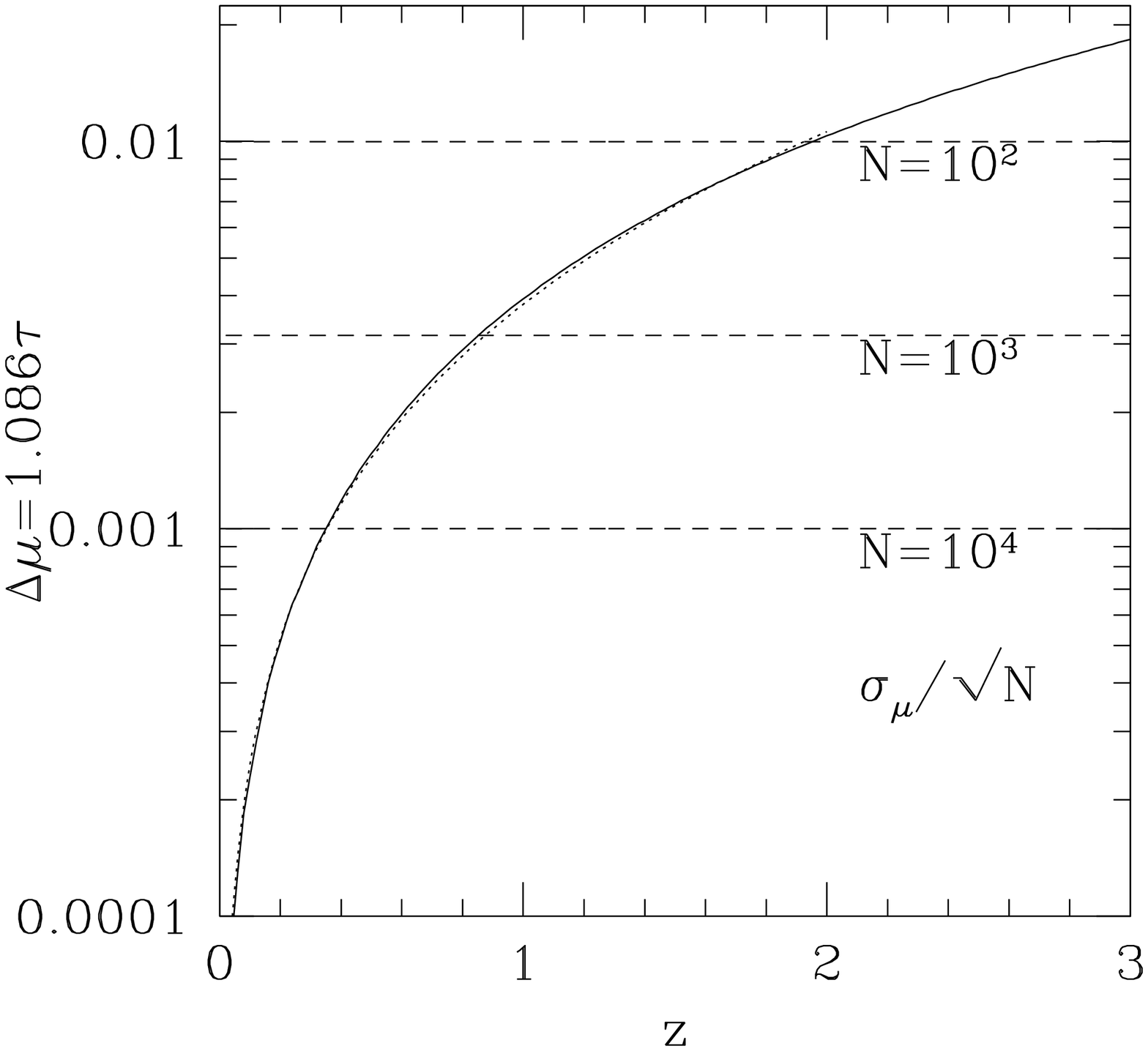}
\caption{Systematic shift in the distance moduli $\mu$ caused by
  Compton scattering (solid line). Although the dimming is only
  $0.4\%$ in flux  at 
  $z=1$ and $1\%$ at $z=2$, the systematical errors induced are
  comparable to statistical errors induced by intrinsic dispersion in
  SN fluxes for future SN surveys  with $\sim 1000$ SNe Ia at
  $z>1$. Here intrinsic dispersion $\sigma_{\mu}=0.1$ mag is adopted
  and the associated statistical errors are shown as dash
  lines. The function form  $\mu^L z+\mu^Qz^2$ adopted to handle
  possible unknown systematic errors is an excellent
  parametrization for this type of $\Delta \mu$ at $z\leq 2$, which is
  shown as the dot  line, almost indistinguishable from the real
  $\Delta \mu$. \label{fig:mu}}
\efi

$\tau$ increases quickly with redshifts, scales as $(1+z)^3-1$ at low
redshift and $(1+z)^{3/2}$ at high redshift until the epoch of
reionization. We adopt the WMAP cosmology 
with $\Omega_m=0.26$, $\Omega_{\Lambda}=1-\Omega_m$ and 
$\Omega_bh=0.032$ \citep{WMAP3} for the numerical evaluation.  Compton
scattering dims the supernova flux by $0.004$ mag at $z=1$ and 
$0.01$ mag at $z=2$ (Fig. \ref{fig:mu}). This dimming is far from
sufficient to challenge the existence of dark energy. Nonetheless, its
impact is non-negligible for precision cosmology based on
supernovae. Future SN surveys such as  JDEM, will measure $\sim
1000$ SNe Ia at $z>1$. For these surveys, the major statistical
uncertainty is the SN intrinsic fluctuations. With $N\sim 1000$ SNe,
 intrinsic fluctuations are reduced to a level of
$\sigma_{\mu}/\sqrt{N}\simeq 0.003$ mag. Here $\sigma_{\mu}$ is the intrinsic
dispersion in SN luminosities. So the Compton dimming  must be
corrected, otherwise the induced systematical errors would be
comparable to the statistical errors. LSST can measure $\sim 10^5$ SNe Ia to $z\sim 1$. For it, there are extra errors 
associated with photo-z uncertainties, whose dispersion is $\sim
0.01$-$0.1$. Even so,  for LSST, systematical error induced by Compton
scattering likely overwhelms the statistical errors induced by
intrinsic fluctuations and photo-z uncertainties.

Since the Compton dimming increases toward high redshift, it biases
$w$ toward more negative value at higher redshift
(Fig. \ref{fig:w}). We isolate the impact on $w$ by fixing other cosmological 
parameters  at their fiducial values and show the result in
Fig. \ref{fig:w}. The systematic shift $\Delta w\sim -2\tau$ (roughly
$4$  times the fractional error in the distance). It  shifts $w$ by
$-0.008$ for SNe Ia at $z=1$ and by 
$-0.017$ for SNe Ia at $z=1.7$. This systematic bias is comparable to
the rms  uncertainty in the pivot $w$ for stage IV SN 
surveys \citep{DETF}. Thus it is of particular importance for the key
dark energy task, to  confirm or invalidate the existence of the cosmological constant.

\bfi{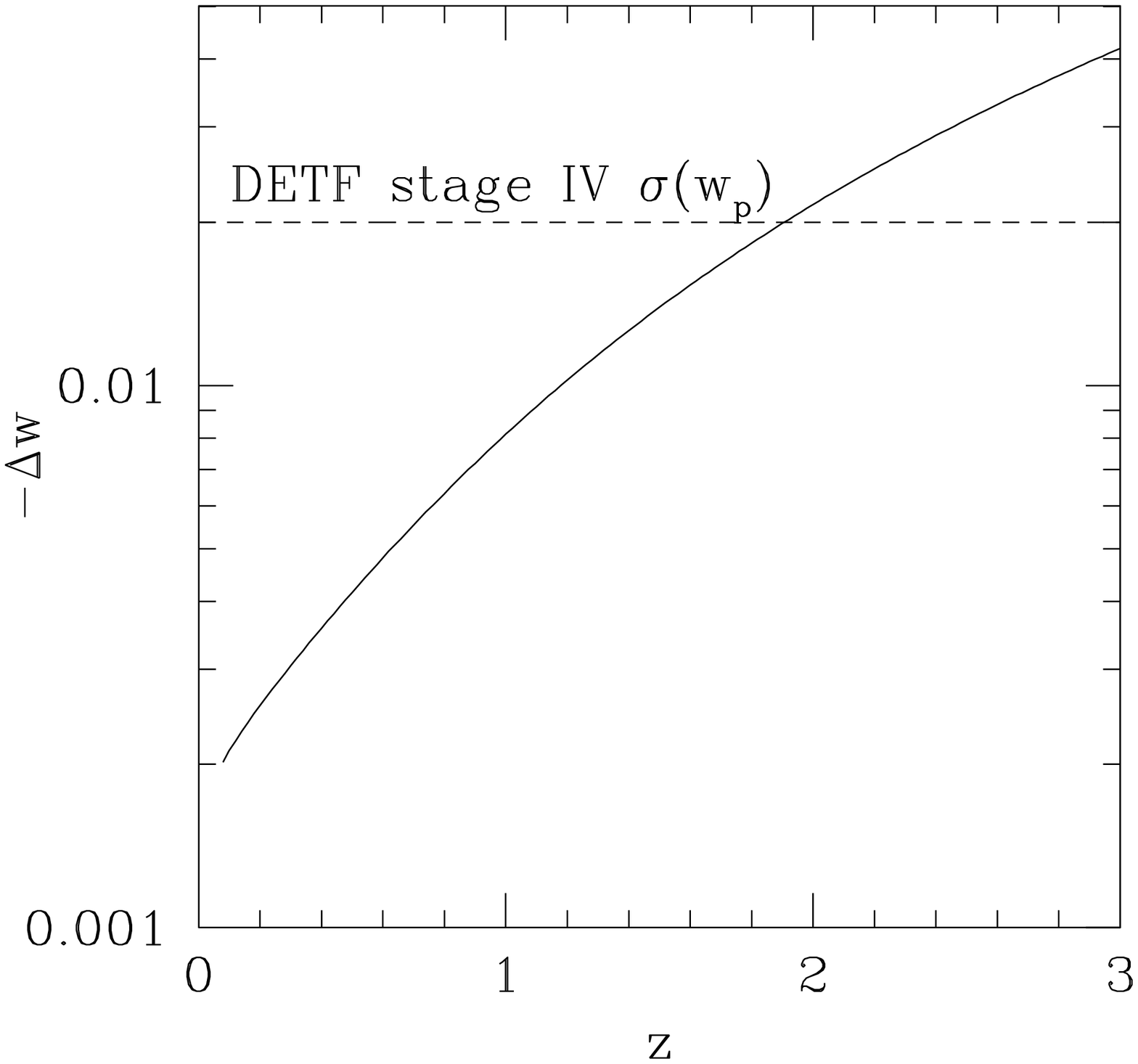}
\caption{Bias in $w$ induced by the Compton dimming (solid line). In this
  estimation, other cosmological parameters are fixed at their
  fiducial values in order to isolate the impact on $w$. Roughly $\Delta w\sim -2\Delta
  \mu\simeq -2\tau$. The dash line is the rms uncertainty in $w=w_p$ at the
  pivot $a=a_p$ for the stage IV supernova surveys \citep{DETF}.  \label{fig:w}}
\efi

Clearly we must take this effect into the analysis of these future
surveys in order not to bias the cosmological
constraints. Observationally, this effect can not be corrected, since
it lacks observational consequences such as reddening that can be
applied to separate from other effects. Can  commonly
adopted parametrization of systematic uncertainties such as the
intrinsic evolution well incorporate
this effect? The answer is yes. It can be fitted with excellent
accuracy by $\Delta \mu=\mu^Lz+\mu^Qz^2$ with $\mu^L=2.3\times 10^{-3}$ and
$\mu^Q=1.5\times 10^{-3}$ (Fig. \ref{fig:mu}).  On one hand, this
means that, the Compton dimming is automatically corrected through
this kind of self calibration. On the other hand, this implies that,
without knowledge of  the Compton dimming, it could be
misinterpreted as an intrinsic  evolution in supernova luminosity. 

 Fortunately, this effect is straightforward to take into 
account from the theory part. Besides the cosmological parameter
$\Omega_m$, the dark energy density $\Omega_{DE}$ and equation of
state $w$ that supernova cosmology aims to
constrain, only an extra input of  $\Omega_b
h$ ($\tau\propto \Omega_bh$) is required to predict
$\tau$. Furthermore, we do not need the exact number of 
$\Omega_bh$ to perform this correction.  $10\%$ accuracy in
$\Omega_bh$ is sufficient 
to render this source of error negligible for any foreseeable
surveys. Current constraint from CMB already reaches this
accuracy. So, the Compton dimming is completely  correctable.

So far we have implicitly neglected fluctuations in $\tau$ along
different lines of sight, so $\tau$ calculated above is actually the ensemble
average $\langle \tau\rangle$.  In reality, there are fluctuations
  in $\tau$ along different lines of sight. For SNe Ia, which are observed at $z\la
2$, the ionization fraction $X_e(z)=1$ is an excellent
approximation. Fluctuations in $\tau$ are thus mainly caused by
fluctuations in the electron number density.     It is straightforward to show
that $\sigma_{\tau}/\langle \tau\rangle \ll 1$, where $\sigma_{\tau}$
is the rms fluctuation in $\tau$. Since the Compton dimming is already
a small effect, tiny fluctuations above its mean do not cause any
observable effect and can thus be safely neglected. 
\bfi{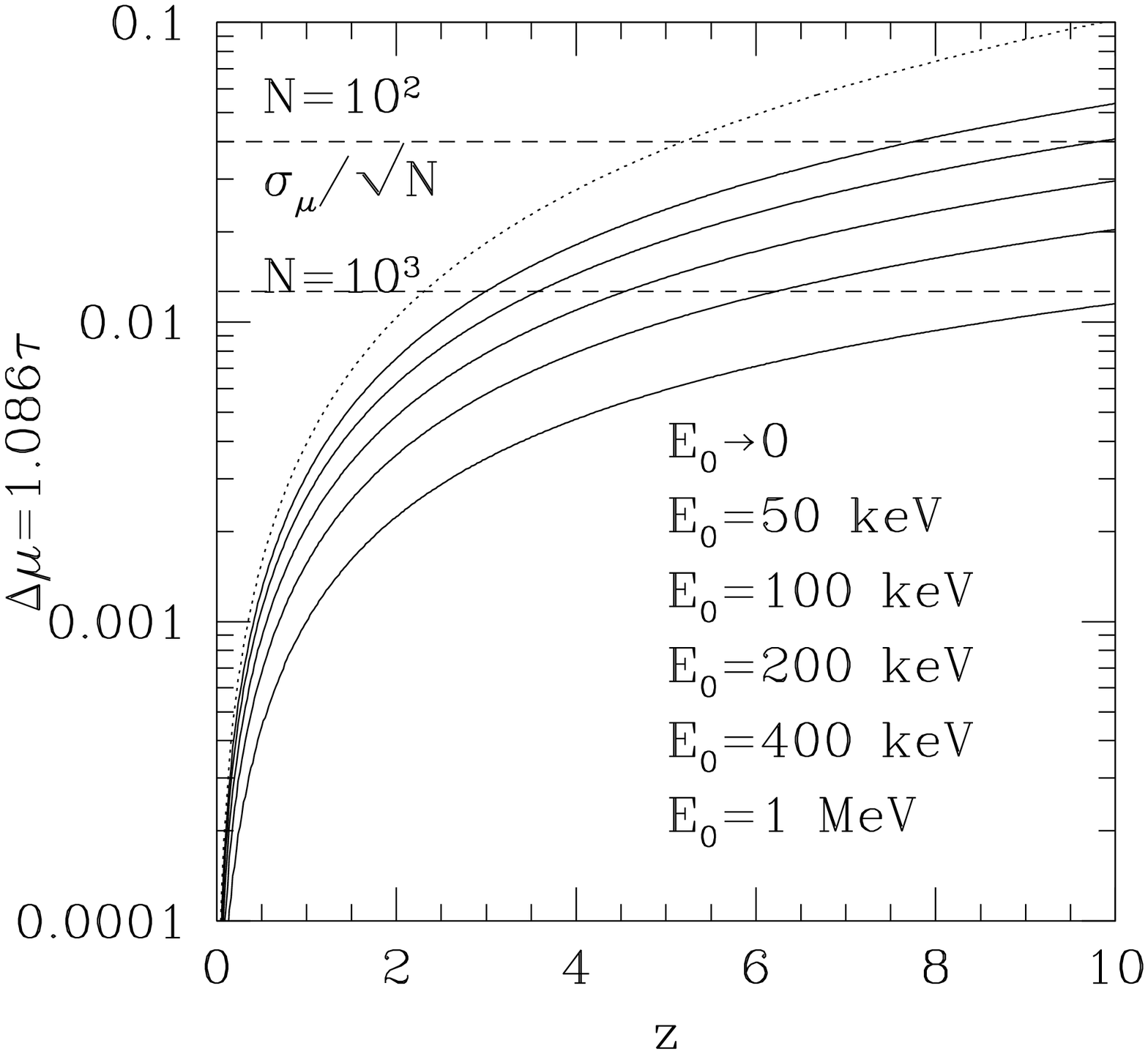}
\caption{The systematic shift in the distance moduli $\Delta \mu$ for
  GRBs, assuming the universe is completely ionized at $z\leq 10$. Since $\gamma$-ray photons are energetic, the Compton
  scattering cross section is now energy dependent. This causes
  $\Delta \mu$ and $\tau$ to decrease with photon energy. Here, $E_0$ is the
  redshifted $\gamma$-ray photon energy. The dash lines are the
  statistical errors for $100$ and $1000$ GRBs,
  respectively. $\sigma_{\mu}=0.4$ mag is adopted for GRBs. \label{fig:GRB}}
\efi
\section{Compton dimming of Gamma ray bursts}
Gamma ray bursts (GRB) are likely standardizable and can serve as
cosmological standard candles (\citet{Xu05} and references therein). They 
have the merit to be sufficiently bright to be observed at redshift
$z>6$ and thus provide important cosmological constraints complementary to SNe Ia.   

GRBs suffer similar Compton dimming. But
since the $\gamma$ photon energy is now comparable to the electron
mass, the cross section of the Compton scattering is
suppressed and becomes energy dependent. $\sigma_T$ in
Eq. \ref{eqn:tau} should be replaced by 
$\sigma(E_0(1+z)/m_ec^2)$, which is given by the Klein-Nishina formula
$\sigma(x)=\frac{3}{4}\sigma_T\{(1+x)[2x(1+x)/(1+2x)-\ln(1+2x)]/x^3+\ln(1+2x)/(2x)-(1+3x)/(1+2x)^2\}$
\citep{Rybicki76}. Here 
$E_0$ is the observed redshifted energy of $\gamma$-ray photons. $\tau$ and
$\Delta \mu$ are now energy (frequency) dependent. The Compton dimming
decreases with the photon energy (Fig. \ref{fig:GRB}). Despite the
suppression in cross section, the Compton dimming can still reach $\sim
0.01$-$0.05$ mag, because GRBs often reside at high redshifts. GRBs
have  larger  intrinsic fluctuations than SNe Ia. However, if more
than a hundred high   redshift GRBs are observed and applied to
constrain cosmology,  this Compton dimming may become non-negligible.

Similar to the case of SNe Ia, it is straightforward to correct
  this Compton dimming effect for GRBs at $z<6$. However, correcting
  it for GRBs at $z>6$ is subtle. For example, patchy reionization can
  cause order of unity 
  fluctuations in $\tau$ along different lines of sight, as suggested
  in radiative transfer  simulations \citep{Holder07}. This could forbid
  complete correction of Compton dimming from the theory part, even
  if the average reionization fraction $X_e(z)$ at $z>6$ is perfectly
  known. Since  the reionization
  history is poorly 
  understood, we are not able to estimate to what level the Compton dimming
  can be corrected for these high redshift GRBs.  A possibility is to
  rely on other surveys to measure $\tau$ along each line of sight and
  to correct for the Compton dimming. For example, future 21-cm surveys of the
  reionization epoch can be applied to reconstruct the optical depth
  along each line of sight, 
  through tight correlation between the optical depth and the 21-cm
  brightness temperature \citep{Holder07}.

\section{Summary}
We point out  that Compton scattering dims supernova flux at a level
non-negligible for future supernova cosmology and must be
taken into the analysis. It also has similar impact on cosmology based
on GRBs.

\acknowledgments
{\it Acknowledgments}:  I thank the anonymous referee for useful
suggestions. This work is supported   by the one-hundred-talents  
program of the Chinese academy of science (CAS), the national science
foundation of China (grant No. 10533030 \& 10543004), the CAS grant
KJCX3-SYW-N2 and the 973 program grant No. 2007CB815401.

\end{document}